\documentclass[fleqn,12pt]{article}
\usepackage{espcrc1}
\usepackage{graphicx}
\title{Finite formation time effects in  inclusive and semi-inclusive
 electro-disintegration of few-body nuclei\footnote{Presented
  at the {\it Third
International Conference on Perspective in Hadronic Physics}, Trieste,
 7-11 May, 2001. To appear in  {\it  Nuclear Physics}.}}

\author{
   H. Morita\address{Faculty of Social Information,
    Sapporo Gakuin University, 11 Bunkyo-dai, Ebetsu,
     Hokkaido 069-8555, Japan},
    M.A. Braun\address{Department of High-Energy Physics,
     S.Petersburg University, 198904 S.Petersburg, Russia},
   C. Ciofi degli Atti\address{Department of Physics,
    University of Perugia, and Istituto Nazionale di
     Fisica Nucleare,
    Sezione di Perugia, Via A. Pascoli, I-06100 Perugia, Italy},
   D. Treleani\address{Department of Theoretical Physics,
    University of Trieste, Strada Costiera 11, Istituto
    Nazionale di Fisica Nucleare, Sezione di Trieste,
     and ICTP, I-34014 Trieste, Italy}
 }

\begin{document}

\maketitle

\begin{abstract}
Finite Formation Time (FFT) effects in the  exclusive reaction
$^4$He(e,e'p)$^3$H at high values of  $Q^2$ are introduced and discussed.
It is shown that the minimum in the
momentum distributions predicted by the Plane Wave Impulse
Approximation (PWIA), which is filled by the Glauber-type
Final State Interaction (FSI), is completely recovered when
FFT effects are taken into account. The semi-inclusive process $^4$He(e,e'p)X
is also investigated.
\end{abstract}

\section{Introduction}
Recently \cite{Braun00} the Glauber approach to Final State Interaction
(FSI) in the inclusive quasi-elastic $A(e,e'p)X$ process, has been extended
  by taking into account the virtuality of the hit  nucleon
  after $\gamma*$ absorption.
  It has been found that at large $Q^2$, due to the fact that the hit
  nucleon need a Finite Formation Time (FFT) to reach
  its asymptotic form, the interaction
   with the remainder of the target nucleus becomes very  weak and vanishes
    in the asymptotic
   limit. In this contribution, we present the results
   of a calculation based upon the extension of the method of \cite{Braun00}
   to the exclusive,  $^4$He(e,e'p)$^3$H, and semi-inclusive, $^4$He (e,e'p)X
   , processes.
\section{The Distorted momentum distributions }
Under certain approximations whose validity will not
be discussed here (see e.g. \cite{Morita00}, the cross section for the
process $A(e,e'p)X$ , can be shown to have the following form
\begin{eqnarray}
  \frac{d^5\sigma}{d\vec{k}_{e'}d\Omega_{\vec{k}_p}} = \mathcal K n_D(\vec{k}_m), \qquad \vec{k}_m=\vec{q}-\vec{k}_p,
\end{eqnarray}
where $\mathcal K$ is a kinematical factor and $n_D(\vec{k}_m)$
the nucleon  distorted momentum distributions, which
 in the case of the  semi-inclusive process $^4$He(e,e'p)X, are  defined as
 follows
\begin{eqnarray}
n_D(\vec{k}_m) &=& (2\pi)^{-3}\int d\vec{r}d\vec{r}'
              exp( i\vec{k}_m\cdot (\vec{r}-\vec{r}') )
              \rho_D(\vec{r},\vec{r}')
\end{eqnarray}
with
\begin{eqnarray}
  \rho_D(\vec{r},\vec{r}') &=& \int d\vec{R}_1d\vec{R}_2
   \psi_{\alpha}^*(\vec{R}_1,\vec{R}_2,\vec{R}_3=\vec{r})S_G^{\dagger }S_G'
   \psi_{\alpha}(\vec{R}_1,\vec{R}_2,\vec{R}_3'=\vec{r}').
\end{eqnarray}

\noindent being the one-body mixed density matrix. In the above equation
 $\psi_{\alpha}$ denotes the $^4$He wave function, ${\vec R}_i's$ are
 usual Jacobi coordinates
 $\vec{R}_1=\vec{r}_2-\vec{r}_1$, $\vec{R}_2=\vec{r}_3-(\vec{r}_1+\vec{r}_2)/2$,
   $\vec{R}_3=\vec{r}_4-(\vec{r}_1+\vec{r}_2+\vec{r}_3)/3$,
   and  $S_G$ is  the Glauber operator.

In the exclusive process $^4$He(e,e'p)$^3$H, $n_D(\vec{k}_m)$ is given by
\begin{eqnarray}
n_D(\vec{k}_m) &=& |w(\vec{k}_m)|^2
\end{eqnarray}
where
\begin{eqnarray}
w(\vec{k}_m)  &=& (2\pi)^{-3/2}\int d\vec{r}
exp(-i\vec{k}_m\cdot\vec{r})A(\vec{r})
\end{eqnarray}
and
\begin{eqnarray}
A(\vec{r})     &=& \sqrt{4}\int d\vec{R}_1 d\vec{R}_2 d\vec{R}_2
   \psi_{t}^*(\vec{R}_1,\vec{R}_2)S_G^{\dagger }
   \psi_{\alpha}(\vec{R}_1,\vec{R}_2,\vec{R}_3=\vec{r}),
\end{eqnarray}

\noindent with $\psi_t$ denoting the  $^3$H wave function.
In our calculations both the $^4He$ and $^3H$ wave functions,
 which correspond to the  Reid Soft Core V$_8$ interaction,
have been taken from \cite{Morita87}.

\section{The Glauber operator and  the Final State Interaction}
The Glauber operator $S_G$ is explicitly given by \cite{Morita00}
\begin{eqnarray}
S_G   = \prod_{i=1}^{3} G(4i),  \qquad
G(4i) = 1-\theta(z_i-z_4)\Gamma(\vec{b}_4-\vec{b}_i),
\label{eq:Glauber}
\end{eqnarray}
where the knocked out nucleon is denoted by "4". Here the $z$-axis is oriented
along  the direction of the motion of the knocked out proton, $\vec{b}$ is
the  component of the nucleon  coordinate in the $xy$ plane, and $\Gamma$
stands for the  usual Glauber profile function
\begin{eqnarray}
\Gamma(\vec{b}) = \frac{\sigma_{tot}(1-i\alpha)}{4\pi b_0^2} e^{-\vec{b}^2/2b_0^2},
\end{eqnarray}
with $\sigma_{tot}$ denoting  the total proton-nucleon cross section,
and $\alpha$ the ratio of the real to imaginary parts of the forward
elastic $pN$ scattering amplitude. The value of $\sigma_{tot}$ and $\alpha$
were chosen at the proper values of the invariant mass of the process,
and the numerical values were taken from
 ref. \cite{PDG}, whereas the value of   $b_0$ has been
  determined by using the relation
  $ \sigma_{el}=\sigma_{tot}^2 (1+\alpha^2) / 16\pi b_0^2$.

When FFT effects are considered, the $G(4i)$ in eq. (\ref{eq:Glauber})
 is replaced by \cite{Braun00}
\begin{eqnarray}
G(4i)       = 1-\mathcal{J}(z_i-z_4)\Gamma(\vec{b}_4-\vec{b}_i), \qquad
\mathcal{J}(z) = \theta(z)(1-exp(\frac{zxmM^2}{Q^2})),
\label{eq:FFT}
\end{eqnarray}
where $x$ is  the Bjorken scaling  variable, $m$ the  nucleon mass, and  $M$
 represents the  average virtuality defined by $M^2 = (m^*_{Av})^2 - m^2$.
 Eq. (9) shows that at high values of  $Q^2$
 FFT effects reduce the  Glauber-type FSI, depending on the value of
 $M$.
In our  calculations the value of the average excitation
mass  $m^*_{Av}$ was  taken to be $1.8(GeV/c)$ \cite{Braun00}.

\section{Results of calculations}
\begin{figure}[htbp]
      \includegraphics[width=75mm]{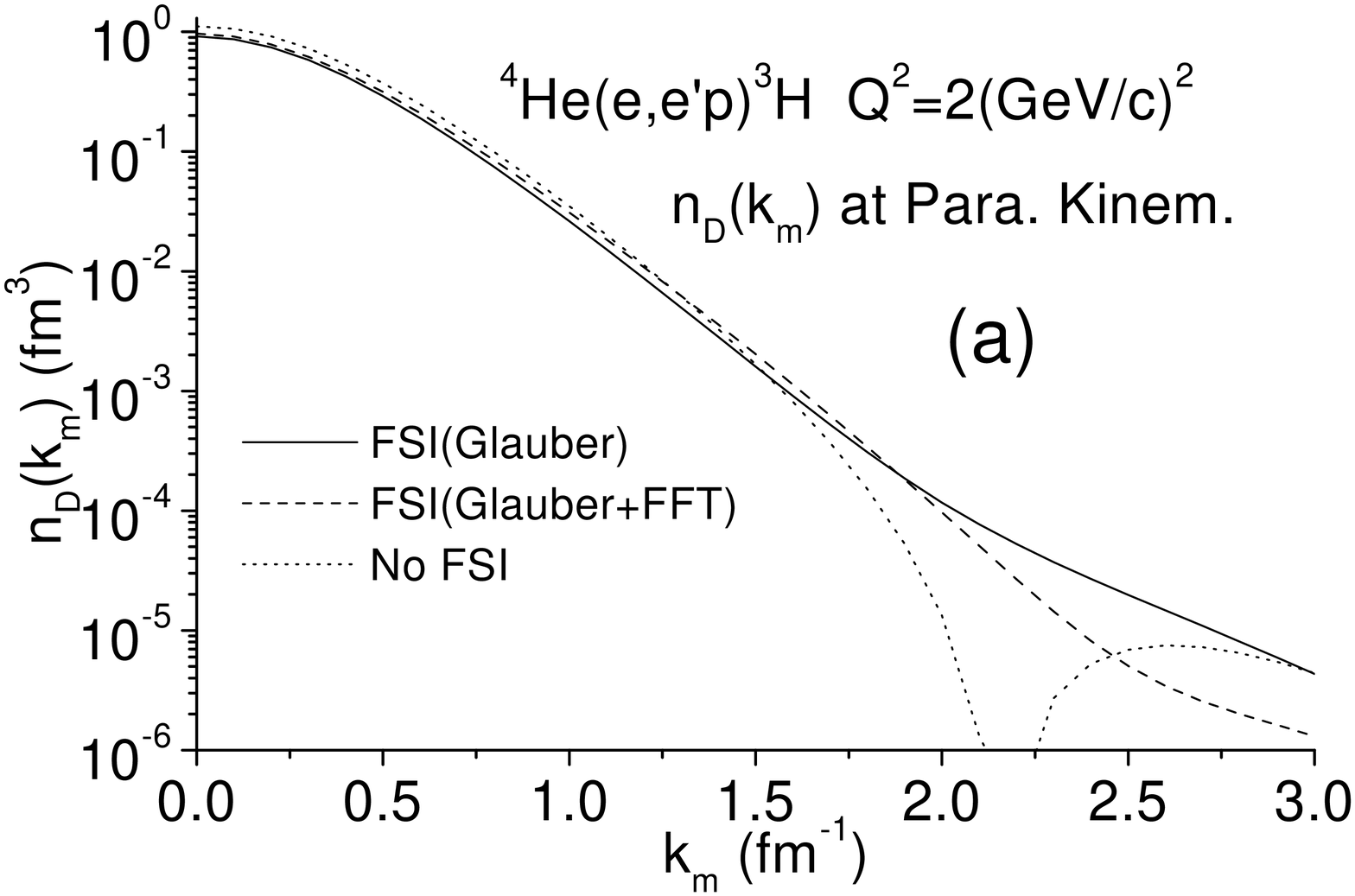}
  \hspace{\fill}
      \includegraphics[width=75mm]{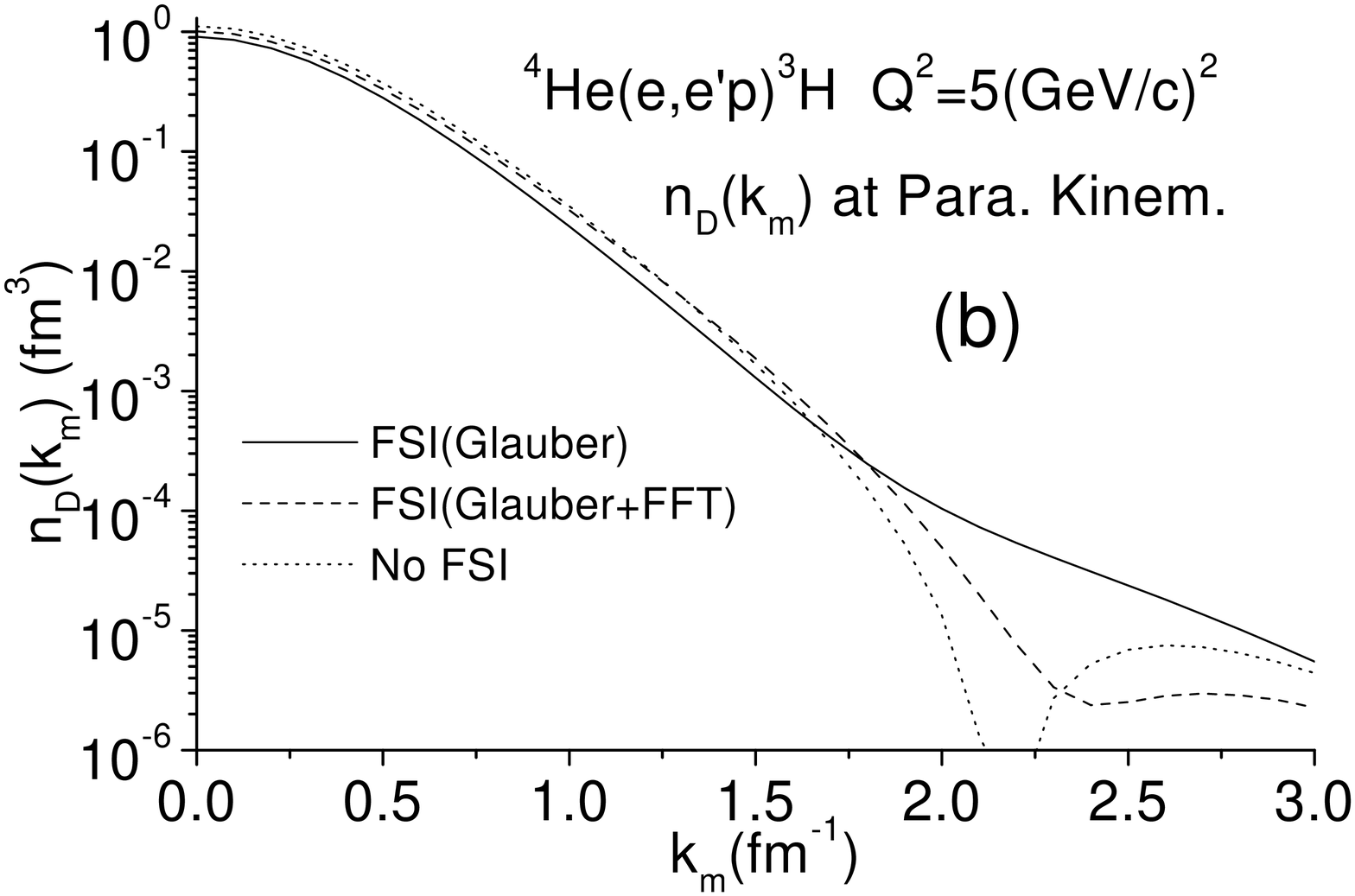}
      \includegraphics[width=75mm]{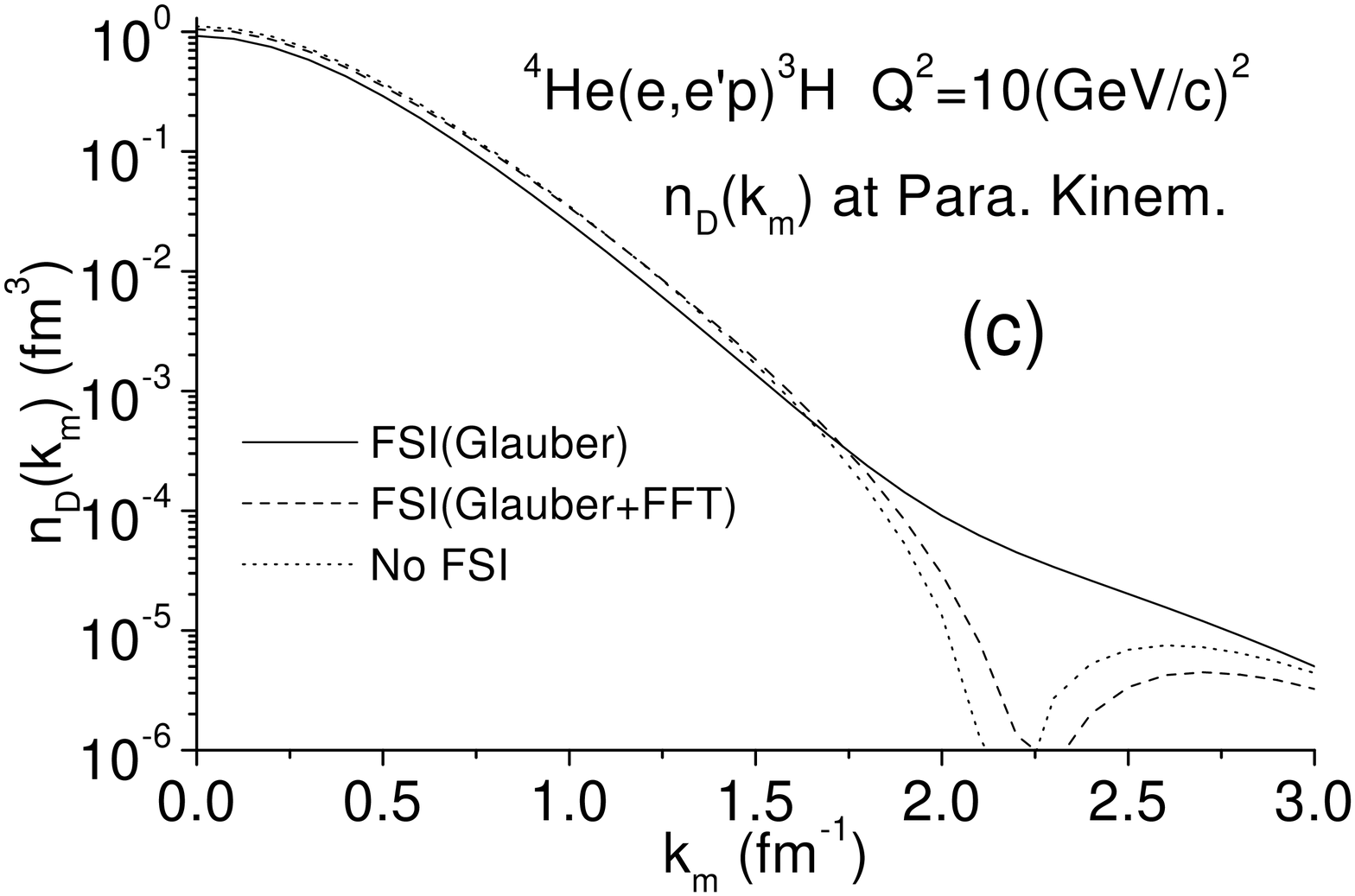}
  \hspace{\fill}
      \includegraphics[width=75mm]{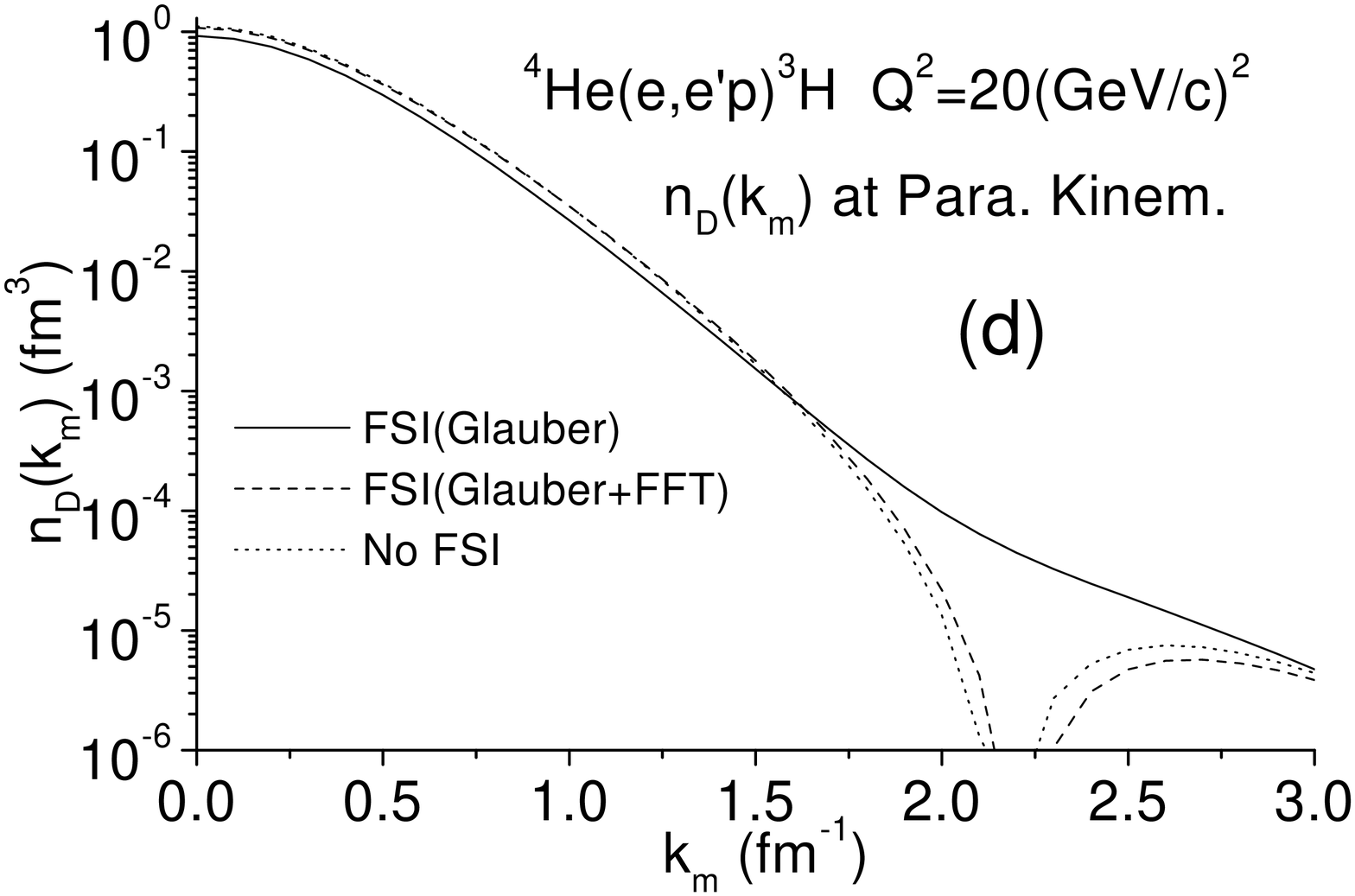}
      \caption{$Q^2$ dependence of $n_D(\vec{k}_m)$.
      Dotted curve: undistorted $n(k_m)$; full curve:
      Glauber-type FSI; dashed curve: Glauber-type FSI+FFT.
        (a) $\sim$ (d) correspond to $Q^2$=2,5,10,20 $(GeV/c)^2$, respectively.}
      \label{fig:fig1}
\end{figure}

The distorted momentum distributions $n_D(\vec{k}_m)$ for the exclusive process
 $^4$He(e,e'p)$^3$H reaction are shown in Fig. \ref{fig:fig1} versus
 the missing momentum
 $\vec{k}_m$, in correspondence of the parallel kinematics,
 i.e. when the missing momentum
   $\vec{k}_m$ is oriented along the virtual photon momentum $\vec{q}$.

Concerning the results presented in Fig. 1, the following remarks are in order.
The  PWIA predicts a minimum at
 $k \sim 2.2 (fm^{-1})$, which is completely filled up by the Glauber-type FSI. The latter
 exhibits a very small $Q^2$ dependence, unlike FFT effects which depend on $Q^2$
 in such a way that  at  $Q^2=20(GeV/c)^2$  they completely
 kill the FSI, so that the minimum is fully recovered.

\begin{figure}[htbp]
\begin{center}
      \includegraphics[width=85mm]{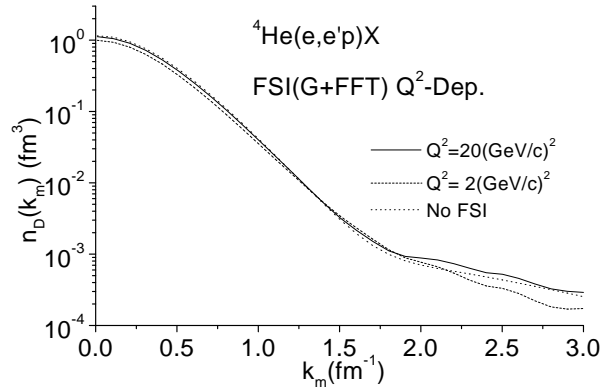}
\end{center}
      \caption{The $Q^2$ dependence of $n_D(\vec{k})$ in the semi-inclusive
       $^4$He(e,e'p)X process.
       Dotted curve: undistorted $n(k_m)$; dashed (full) curve: Glauber-type FSI+FFT
       at $Q^2=2(20) (GeV/c)^2$.}
      \label{fig:fig2}
\end{figure}

 The FFT effects on the semi-inclusive reaction   $^4$He(e,e'p)X  at parallel kinematics
 are shown in  Fig. \ref{fig:fig2}, and it can be seen that here the effects from
 Glauber-type FSI and FFT, are less relevant.

\section{Summary}
The results of the calculations that we have exhibited, show
that the exclusive process $^4$He(e,e'p)$^3$H at high values of $Q^2$
could provide a clear cut check of various models which go beyond the
treatment of FSI effects in terms of Glauber-type rescattering.
In particular, the FFT approach of Ref. \cite{Braun00} predicts a
clean and regular $Q^2$ behaviour leading to the vanishing of FSI effects
 at moderately large values of $Q^2$; such a prediction would be
 validated by the experimental observation of a dip in the cross
 section at $k_m \simeq 2.2 fm^{-1}$. Recently, Benhar {\it et al}
 \cite{Benhar00} have analyzed the same process we have considered
 , viz. the  $^4$He(e,e'p)$^3$H reaction, using a colour transparency model.
 At variance with our results, their model does not lead to the vanishing of
  FSI at  $Q^2 \simeq 20(GeV/c)^2$. We conclude, therefore, that exclusive electron
  scattering off $^4He$ would really represent a powerful tool to
  discriminate various models of hadronic final state rescattering.

\end{document}